\documentclass[preprint,showpacs,preprintnumbers,amsmath,amssymb]{revtex4}
\usepackage{booktabs}
\usepackage{mathrsfs}
\usepackage{epsfig}
\usepackage{graphicx}
\usepackage{dcolumn}
\usepackage{bm}
\usepackage{amsmath}

\let\jnfont=\rm
\def\NPB#1,{{\jnfont Nucl.\ Phys.\ B }{\bf #1},}
\def\PLB#1,{{\jnfont Phys.\ Lett.\ B }{\bf #1},}
\def\EPJC#1,{{\jnfont Eur.\ Phys.\ Jour.\ C }{\bf #1},}
\def\PRD#1,{{\jnfont Phys.\ Rev.\ D }{\bf #1},}
\def\PRL#1,{{\jnfont Phys.\ Rev.\ Lett.\ }{\bf #1},}
\def\MPLA#1,{{\jnfont Mod.\ Phys.\ Lett.\ A }{\bf #1},}
\def\JPG#1,{{\jnfont J.\ Phys.\ G}{\bf #1},}
\def\CTP#1,{{\jnfont Commun.\ Theor.\ Phys.\ }{\bf #1},}
\def\ZPC#1,{{\jnfont Z.\ Phys.\ C }{\bf #1},}
\def\JHEP#1,{{\jnfont JHEP \ }{\bf #1},}
\def\Rv{\not{\hbox{\kern-1pt $R$}}}
\def\p{\not{\hbox{\kern-3pt $p$}}}

\begin{document}
\preprint{\parbox{1.2in}{\noindent arXiv:0912.1447}}

\title{Top quark forward-backward asymmetry at
       the Tevatron: \\ a comparative study in different new physics models}

\author{Junjie Cao$^1$, Zhaoxia Heng$^2$, Lei Wu$^1$, Jin Min Yang$^2$
        \\~ \vspace*{-0.3cm} }
\affiliation{
$^1$ College of Physics and Information Engineering, Henan Normal University,
     Xinxiang 453007, China\\
$^2$ Key Laboratory of Frontiers in Theoretical Physics,
     Institute of Theoretical Physics,
     Academia Sinica, Beijing 100190, China
     \vspace*{1.5cm}}

\begin{abstract}
The top quark forward-backward asymmetry $A_{FB}^{t}$ measured at
the Tevatron is above the Standard Model prediction by more than
2$\sigma$ deviation, which might be a harbinger for new physics. In
this work we examine the contribution to $A_{FB}^{t}$ in two
different new physics models: one is the minimal supersymmetric
model without R-parity (RPV-MSSM) which contributes to $A_{FB}^{t}$
via sparticle-mediated $t$-channel process $d \bar d \to t \bar{t}$;
the other is the third-generation enhanced left-right model (LR
model) which contributes to $A_{FB}^{t}$ via $Z^\prime$-mediated
$t$-channel or $s$-channel processes. We find that in the parameter
space allowed by the $t\bar{t}$ production rate and the $t\bar{t}$
invariant mass distribution at the Tevatron, the LR model can
enhance $A^t_{FB}$ to within the $2\sigma$ region  of the Tevatron
data for the major part of the parameter space, and in optimal case
$A^t_{FB}$ can reach $12\%$ which is slightly below  the $1\sigma$
lower bound. For the RPV-MSSM, only in a narrow part of the
parameter space can the $\lambda''$ couplings enhance $A^t_{FB}$ to
within the $2\sigma$ region while the $\lambda'$ couplings just
produce negative contributions to worsen the fit.

\end{abstract}

\pacs{14.65.Ha,14.80.Ly,11.30.Hv}

\maketitle

\section{INTRODUCTION}
As the heaviest fermion with a mass at weak scale, top quark is
speculated to be a sensitive probe of new physics beyond the
Standard Model (SM) \cite{top-review}. The precision measurement of
its properties, now being performed at the Tevatron and will be
continued at the CERN Large Hadron Collider (LHC), will either
unravel or further constrain the new physics related to the top
quark.

So far the production rates of top pair and single top measured at
the Tevatron are in good agreement with the SM predictions, but a
more than $2\sigma$ deviation is reported in the forward-backward
asymmetry $A_{FB}^{t}$ in top pair production, which is defined by
\begin{eqnarray}
A_{FB}^{p\bar{p}}=\frac{N_t(\cos{\theta}>0)-N_t(\cos{\theta}<0)}
                   {N_t(\cos{\theta}>0)+N_t(\cos{\theta}<0)} ,
\end{eqnarray}
with $\theta$ being the angle between the reconstructed top quark
momentum and the proton beam direction in $t\bar{t}$ rest frame. In
the SM, this asymmetry gets dominant contribution from the
next-to-leading-order QCD correction and
was found to be several percent: ${A_{\mathrm{FB}}^{t}}({\mathrm SM})=5.0\pm
1.5\%$ \cite{top-afb-sm}. Compared with its experimental value
$A_{\mathrm{FB}}^{t}(\mathrm{exp})=19.3\pm 6.9 \%$ measured by the CDF and D0
collaborations \cite{top-afb-exp}, this
SM prediction is below the measured value by more than 2$\sigma$
deviation.

Such a discrepancy might be a new physics footprint in top quark
sector and has been studied in several new physics models, where the
Kaluza-Klein excitations in extra dimensions \cite{top-afb-extra},
the presence of new gauge bosons ($Z'$, $W'$, axigluon)
\cite{top-afb-s,top-afb-t} or new scalars \cite{top-afb-scalar} are
utilized to try to explain the discrepancy. Noting that the
mechanisms proposed in these literatures can also be realized in
some popular new physics models,  we in this work study the
asymmetry $A_{FB}^{t}$ in the supersymmetric models and the
left-right models \cite{LR}.

For the minimal supersymmetric standard model (MSSM) \cite{mssm},
the SUSY influence on $t\bar{t}$ production comes from loop effects,
in which the SUSY-QCD effects are dominant over the SUSY-EW effects
\cite{susytt-loop}. But among the SUSY-QCD one-loop diagrams only
the box diagrams contribute to the asymmetry $A_{FB}^{t}$ and,
consequently, the contribution is negligibly small (see the
discussion in \cite{top-afb-extra}). Therefore, in  our analysis we
consider the R-parity violating MSSM (RPV-MSSM) \cite{rpv} which
allows for tree-level contribution from $t$-channel process $d
\bar{d} \to t \bar{t}$ by exchanging a color-singlet slepton or a
color-triplet squark.  For the general left-right models, since the
predicted new gauge bosons are usually at TeV scale and unlikely to
affect $A^t_{FB}$ significantly, we here consider a special
left-right model called the third-generation enhanced left-right
model \cite{hexg}. This model predicts a new gauge boson $Z'$ which
contributes to $A_{FB}^{t}$ via $t$-channel or $s$-channel
processes.

This paper is organized as follows.  In Sec. II we describe the
calculation of the asymmetry $A^t_{FB}$ in the RPV-MSSM and present
some numerical results and discussions. In Sec. III we
perform similar analysis in the third-generation enhanced left-right
model. Finally, some discussions and the conclusion are presented in
Sec. IV.

\section{$A_{FB}^{t}$ in $R$-parity violating MSSM}
In the popular MSSM, the invariance of $R$-parity, defined by
$R=(-1)^{2S+3B+L}$ for a field with spin $S$, baryon-number $B$ and
lepton-number $L$, is often imposed on the Lagrangian in order to
maintain the separate conservation of $B$ and $L$. Although
$R$-parity plays a beautiful role in the phenomenology of the MSSM
(e.g., forbid proton decay and ensure a perfect candidate for cosmic
dark matter), it is, however, not dictated by any fundamental
principle such as gauge invariance and there is no compelling
theoretical motivation for it. The most general superpotential of
the MSSM consistent with the SM gauge symmetry and supersymmetry
contains $R$-violating interactions which are given by~\cite{rpv}
\begin{equation}\label{poten}
{\cal W}_{\not \! R}=\frac{1}{2}\lambda_{ijk}L_iL_jE_k^c
+\lambda'_{ijk} L_iQ_jD_k^c
+\frac{1}{2}\lambda''_{ijk}\epsilon^{\alpha\beta\gamma}U_{i\alpha}^cD_{j\beta}^cD_{k\gamma}^c
+\mu_iL_iH_2,
\end{equation}
where $i,j,k$ are generation indices, $c$ denotes charge
conjugation, $\alpha$, $\beta$ and $\gamma$ are the color indices with
$\epsilon^{\alpha\beta\gamma}$ being the total antisymmetric tensor,  $H_{2}$ is
the Higgs-doublet chiral superfield, and $L_i(Q_i)$ and
$E_i(U_i,D_i)$ are the left-handed lepton (quark) doublet and
right-handed lepton (quark) singlet chiral superfields. The
dimensionless coefficients $\lambda_{ijk}$ (antisymmetric in $i$ and
$j$) and $\lambda'_{ijk}$
 in the superpotential are
$L$-violating couplings, while $\lambda''_{ijk}$ (antisymmetric in
$j$ and $k$) are $B$-violating couplings.

\begin{figure}[htb]
\epsfig{file=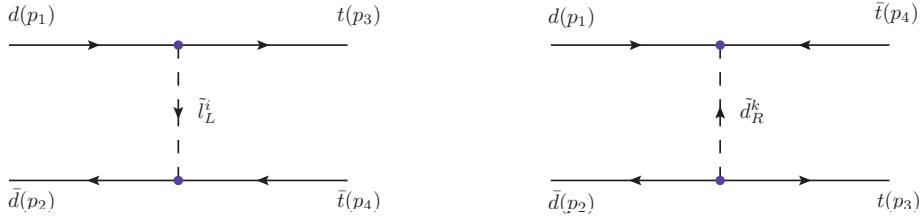,width=13cm} \vspace{-0.5cm} \caption{Feynman
diagrams contributing to $A^t_{FB}$ in the RPV-MSSM, where $\tilde
l^i_L$ and $\tilde d^k_R$ denotes a left-handed slepton in $i$-th
generation and right-handed squark in $k$-th generation,
respectively.} \label{fig1}
\end{figure}

The expression of ${\cal W}_{\not \! R}$ implies that both
$\lambda_{ijk}^\prime$¡¡and $\lambda_{ijk}^{\prime \prime} $ can
induce new top quark interactions. In terms of the four-component
Dirac notation, the interactions involved in top pair production are
\begin{eqnarray}
{\cal L} &=& \lambda^{\prime}_{ijk}
  \tilde l^i_L  \overline{d^k_R} u^j_L - \frac{1}{2}\lambda^{\prime\prime}_{ijk} [ \tilde d^{k*}_R \bar
u^{i}_R d^{jc}_L+\tilde d^{j*}_R\bar u^i_Rd^{kc}_L ]+h.c.,
\end{eqnarray}
and these interactions can contribute to the forward-backward
asymmetry by the diagrams shown in Fig.\ref{fig1}. The corresponding
amplitudes are then given by
\begin{eqnarray}
M_{d\bar{d} \to t \bar{t}}^{RPV}|_{\lambda^\prime} &=&
-i\delta_{\alpha\rho}\delta_{\beta\sigma}|\lambda_{i31}^{'}|^{2}
\frac{\bar{u}(t)P_R
u(d)\bar{v}(d)P_Lv(t)}{(p_1-p_3)^2-m_{\tilde{l}_{iL}}^{2}} , \label{singlet} \\
M_{d\bar{d} \to t \bar{t}}^{RPV}|_{\lambda^{\prime \prime}} &=&
-i\varepsilon_{\beta\rho\lambda}\varepsilon_{\sigma\alpha\lambda}|\lambda_{31k}^{''}|^{2}
\frac{\bar{u}(t)\gamma_\mu P_R v(t)\bar{v}(d)\gamma^\mu
P_Ru(d)}{2[(p_1-p_4)^2-m_{\tilde{d}_{kR}}^{2}]} \label{triplet} ,
\end{eqnarray}
with $\alpha$, $\beta$, $\rho$, $\sigma$ and $\lambda$ being color
indices of the quarks and squarks.  These amplitudes affects
$A^t_{FB}$ by interfering with the QCD amplitude $d\bar{d} \to
g^\ast \to t \bar{t}$ and also by its own square. In our results
presented below, we have included the SM contribution to $A^t_{FB}$
and considered only one coupling non-zero each time.

\begin{table}
\caption{The upper bounds on the couplings $\lambda^{'}_{i31}$ ($i=1,2,3$) and
$\lambda^{''}_{31k}$ ($k=2,3$) \cite{review}. }
 \begin{tabular}{ccl} \hline
 couplings &~~~~~~~~~~~~~~bounds &~~~~~~~~~~~~~~~~~sources\\
 \hline
$\lambda^{\prime}_{131}$ &~~~~~~~~~~~~~ $0.03~m_{\tilde u^i_L}/(100$ GeV)&~~~~~~~~~~~~~~~~~$Q_W(Cs)$\\
$\lambda^{\prime}_{231}$ &~~~~~~~~~~~~~ $0.18~m_{\tilde{d}^k_L}/(100$ GeV)&~~~~~~~~~~~~~~~~~~$\nu_\mu q$ \\
$\lambda^{\prime}_{331}$ &~~~~~~~~~~~~~ $0.26~m_{\tilde{d}^k_R}/(100$ GeV)&~~~~~~~~~~~~~~~~~~$K\to\pi\nu\bar{\nu}$ \\
$\lambda^{''}_{31k}$     &~~~~~~~~~~~~~ $0.97~m_{\tilde d^k_R}/(100$ GeV)  &~~~~~~~~~~~~~~~~~~$R^{Z}_{l}$ \\
$\lambda^{''}_{31k}$     &~~~~~~$1.25$  &~~~~~~~~~~~~~~~~~~perturbativity\\
\hline
 \end{tabular}\label{dlambda}
 \end{table}

The SUSY parameters involved in the calculation are the couplings
$\lambda^\prime_{i31}$ and $\lambda^{''}_{31k}$ as well as sparticle
masses. So far both theorists and experimentalists have intensively
studied the phenomenology of these couplings in various processes
\cite{rp2} and obtained some bounds \cite{review}. In Table
\ref{dlambda} we list the relevant bounds and, as can be seen, in
case of heavy squarks, these bounds are quite weak. Note that for
$\lambda^{''}_{31k}$ we do not use the stringent bound from $n-\bar
n$ oscillation \cite{review} because they depend on additional SUSY
parameters which are not involved in our processes.

\begin{figure}[htbp]
\includegraphics[width=7cm]{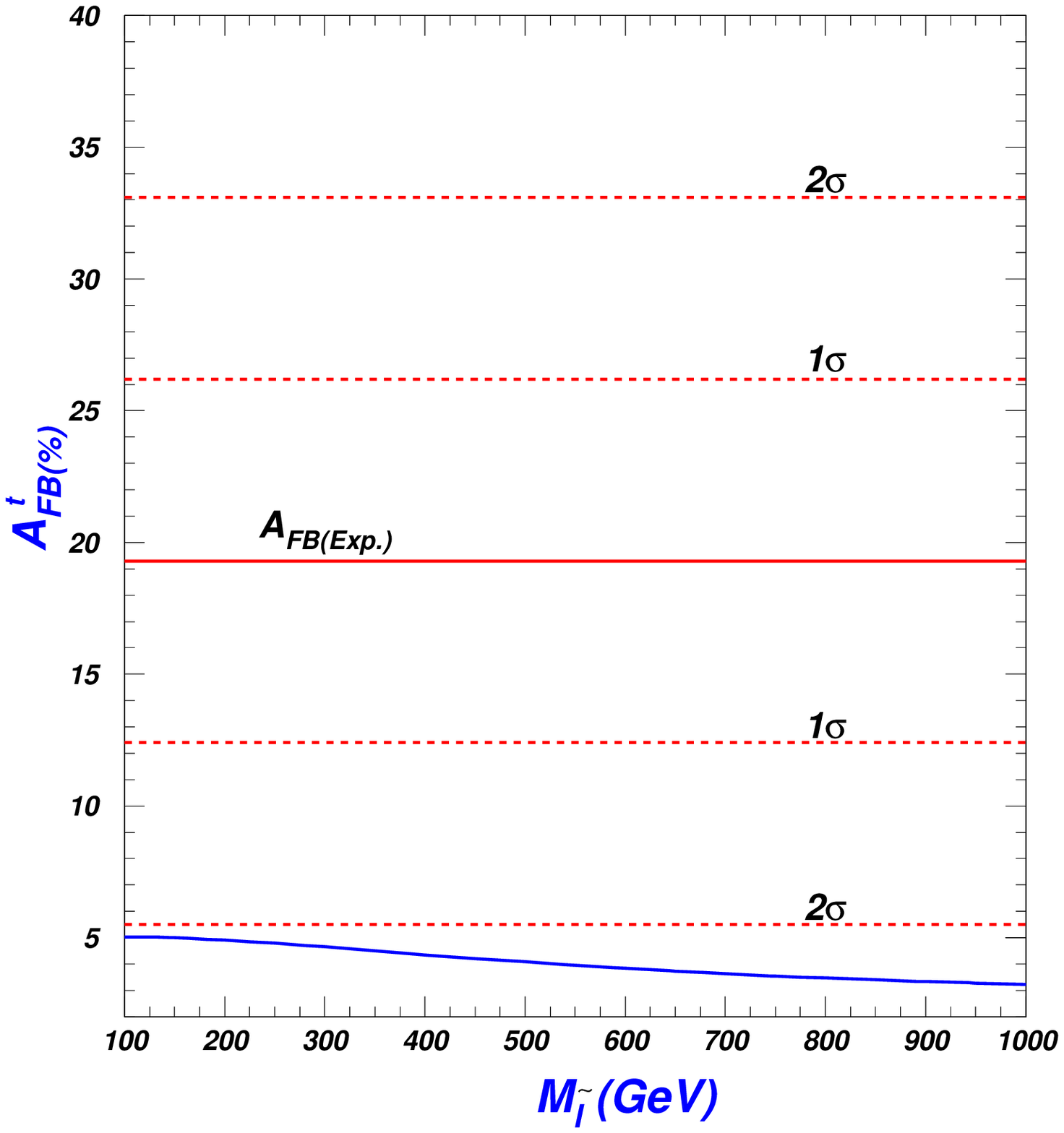}%
\hspace{0in}%
\includegraphics[width=7cm]{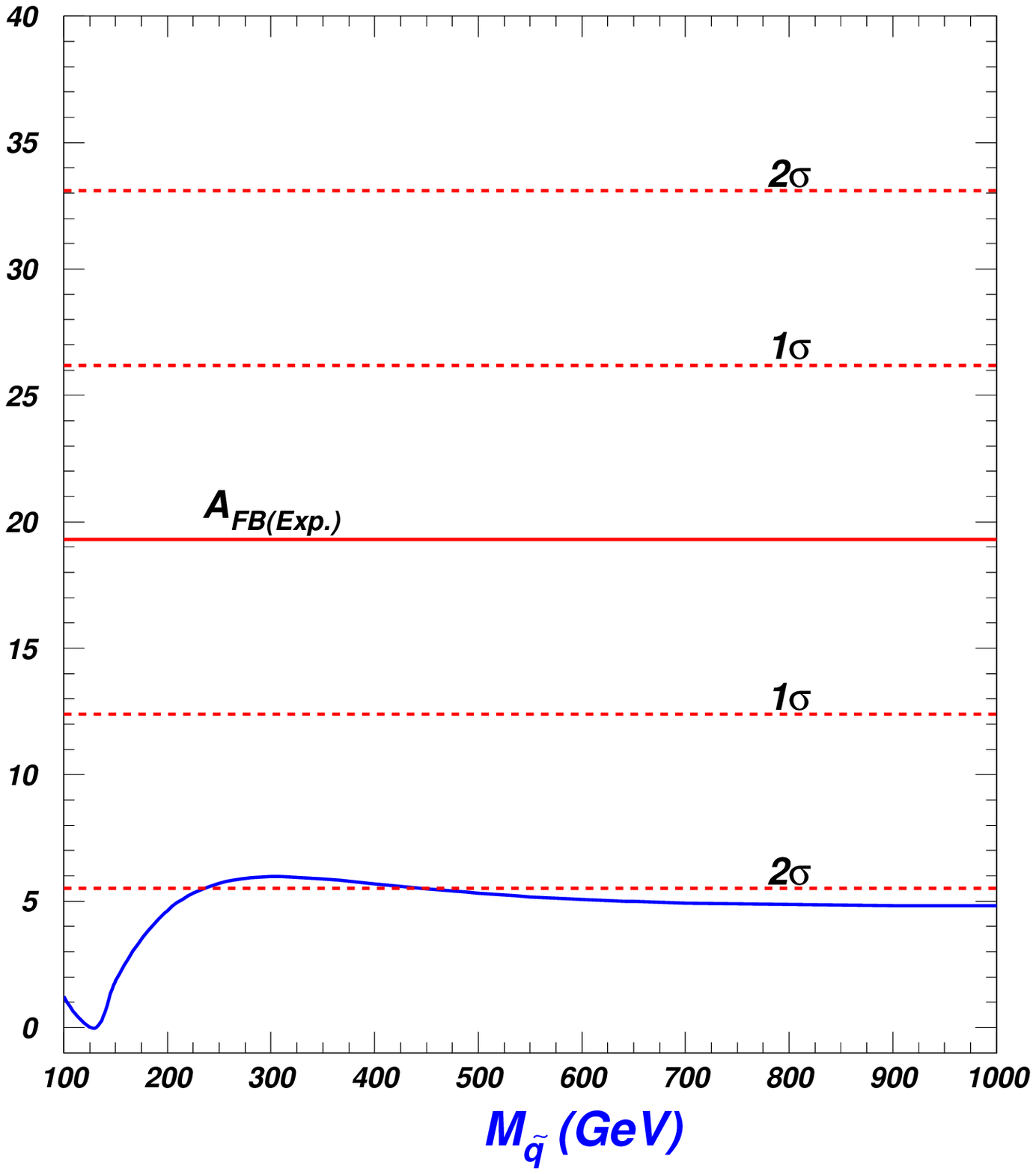}%
\vspace{-0.8cm} \caption{$A_{FB}^{t}$ versus the sparticle mass in
the R-parity violating MSSM. The left (right) panel is induced by
$\lambda^\prime_{i31}$ ($\lambda^{''}_{31k}$) which are fixed at
their maximally allowed values. The SM contribution to $A^t_{FB}$ is
also included.} \label{fig2}
\end{figure}
In Fig.\ref{fig2} we show the dependence of $A^t_{FB}$ on the
relevant sparticle masses. Here we assume all  squarks and sleptons
are degenerate and sum over the contributions from different
generations. The couplings $\lambda^\prime_{i31}$ and
$\lambda^{''}_{31k}$ are fixed at their maximally allowed values
which, as shown in Table I, vary with squark mass. The SM parameters
are taken as \cite{pdg}
\begin{eqnarray}
m_t=172.5{\rm ~GeV},~m_{Z}=91.19 {\rm ~GeV},~\sin\theta_W=0.2228,
~\alpha_s(m_t)=0.1095,~\alpha=1/128.
\end{eqnarray}

Fig.\ref{fig2} indicates that both $\lambda^\prime_{i31}$ and
$\lambda^{\prime \prime}_{31k}$ can give a negative contribution to
$A_{FB}^{t}$ and in the worst cases, they decrease $A^t_{FB}$ by
$2\%$ and $5\%$ respectively, which are comparable in size with
$A^{t\ SM}_{FB}$. In a narrow part of the squark mass
($250 \sim 400$ GeV), the coupling $\lambda^{''}_{31k}$ can also give a
positive contribution to enhance $A_{FB}^{t}$ to within the $2\sigma$
region of the Tevatron data.

We note that the scalar-mediated contributions have been analyzed in
a model independent way in \cite{top-afb-scalar} and the results
were presented for color-singlet, -triplet, -sextet and -octet
scalar respectively. We checked that our analytic results are in
agreement with \cite{top-afb-scalar} for the color-singlet and
-triplet cases except that our study is restricted in a specified
model, the RPV-MSSM.

Before ending this section, we give a comment on $A^t_{FB}$ in the
top-color assisted technicolor model (TC2)\cite{tc2}. This model
predicts some composite bosons, $\pi^{0}_{t}$ and $\pi^{\pm}_{t}$,
with mass at weak scale and having large Yukawa couplings to top
quark. As a result, $A^t_{FB}$ gets additional contribution from
$t$-channel processes $u\bar{u} (c \bar{c}) \to t \bar{t}$ by
exchanging a color-singlet scalar $\pi_t^0$, which is similar to
Fig.\ref{fig1} (a) in RPV-MSSM. Noting the up quark content in
proton is larger than the down quark content, one may expect a
larger effect on $A^t_{FB}$ in TC2 model than in RPV-MSSM. This is
not true because for $u \bar{u} \to t \bar{t}$ in TC2 model, its
contribution is proportional to $u_R-t_R$ mixing which is determined
by the triangular texture of the up-type quark mass matrix and turns
out to be very small \cite{cpyuan}. As to $c \bar{c} \to t \bar{t}$, although
$c_R-t_R$ mixing may be sizable \cite{cpyuan}, it actually gives no
contribution to $A^t_{FB}$ because the distributions of $c$ and
$\bar{c}$ in proton are approximately same and so the initial state
of this process is symmetric under the exchange of $c$ and
$\bar{c}$. In fact, we numerically calculated all TC2 contributions
to $\bar{t} t$ production at the Tevatron, which include the
$s$-channel processes $g g, b \bar{b} \to \pi_t^{0 \ast} \to t
\bar{t}$, $t$-channel processes $b \bar{b} \to t \bar{t}$ by
exchanging $\pi_t^-$ and $u \bar{u}, c \bar{c} \to t \bar{t}$ by
exchanging $\pi_t^0$, and we found that TC2 model can change
$\sigma_{t\bar{t}}$ by at most 200 fb and $A^t_{FB}$ by order of
$10^{-4}$.

\section{ $A_{FB}^{t}$ in the third-generation enhanced left-right model}
In the third-generation enhanced left-right model
\cite{hexg}, the gauge group is $SU(3)_C \times SU(2)_L\times
SU(2)_R \times U(1)_{B-L}$ with gauge couplings $g_3$, $g_L$, $g_R$
and $g$, respectively. This model differs from other left-right
models by having right-handed gauge bosons  couple predominantly to
the third generation fermions. Noting that the right-handed
 gauge boson $Z_R$ will mix with the standard $Z_0$  to form mass
 eigenstates $Z$ and $Z^\prime$, one can
write down the neutral gauge interactions of quarks as \small
\begin{eqnarray}
{\cal L}_Z &=& -{g_L\over 2 \cos\theta_W} \bar q \gamma^\mu (g_V -
g_A \gamma_5) q (\cos\xi_Z Z_\mu - \sin\xi_Z Z^\prime_\mu)
\nonumber\\
&+& {g_Y\over 2} \tan\theta_R ({1\over 3} \bar q_L \gamma^\mu q_L+
{4\over 3} \bar u_{Ri} \gamma^\mu u_{Ri} -{2\over 3} \bar
d_{Ri}\gamma^\mu d_{Ri})
(\sin\xi_Z Z_\mu + \cos\xi_Z Z^\prime_\mu)\nonumber\\
&-& {g_Y\over 2} (\tan\theta_R + \cot\theta_R) ( \bar u_{Ri}
\gamma^\mu V^{u*}_{Rti} V^{u}_{Rtj}u_{Rj} - \bar d_{Ri} \gamma^\mu
V^{d*}_{Rbi} V^{d}_{Rbj} d_{Rj}) (\sin\xi_Z Z_\mu + \cos\xi_Z
Z^\prime_\mu) \label{neucoup}
\end{eqnarray}
\normalsize where $\tan \theta_R = g/g_R$, $g_Y = g \cos\theta_R =
g_R \sin\theta_R$, $\xi_{Z}$ is the mixing angle between $Z_R$ and
$Z_0$, and $V^{u,d}_{Rij}$ are unitary matrices which rotate the
right-handed quarks $u_{Ri}$ and $d_{Ri}$ from interaction basis to
mass eigenstates. This model also predicts new charged gauge
interactions of quarks,  but since their effect on $A^t_{FB}$ is
small, we do not consider them here. Note that in Eq.(\ref{neucoup}),
$q$ and $q_L$ are summed over all quarks, and the repeated generation
indices $i$ and $j$ are also summed.

\begin{figure}[htbp]
\epsfig{file=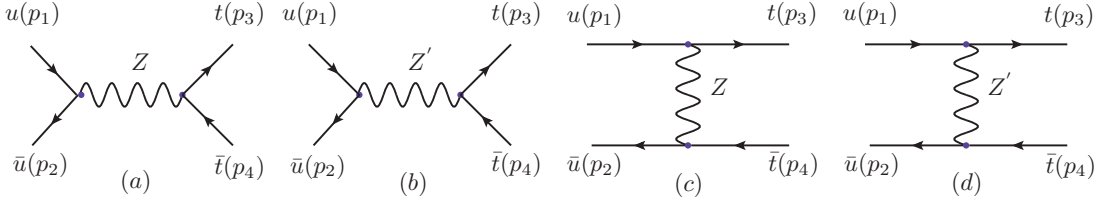,width=15cm} \vspace*{-0.7cm} \caption{Feynman
diagrams contributing to $A^t_{FB}$ in the third-generation enhanced
left-right model.} \label{fig3}
\end{figure}

Eq.(\ref{neucoup}) indicates that the $Z^\prime \bar{u}_i u_j$
interaction is large when $g_R \gg g_Y$ or $\cot \theta_R \gg 1$.
This feature can be utilized to enhance $A^t_{FB}$ by the
$t$-channel process $u \bar{u} \to t \bar{t}$ if $(V^u_R)_{ut}$ is
moderately large. In \cite{hexg}, $V^d_{R}$ and $V^u_{R}$ are
assumed to be nearly diagonal to satisfy the severe flavor-changing
neutral-current (FCNC) constraints. However, as pointed out in
\cite{top-afb-t}, a sizable $u_R-t_R$ mixing does not conflict with
these constraints given that the other flavor mixings are
suppressed. Here we further point out that this pattern of flavor
mixings does not necessarily require the up-top element in up-type
quark mass matrix $M_u$ to be much larger than other
off-diagonal elements. For example, assuming $(V_{R}^u)_{ut} = 0.2$,
$(V_{R}^u)_{ct} =0 $ and $(V_{R}^u)_{uc}=0$, we numerically solve
the equation $V^{u \dagger}_R M_u^\dagger M_u V^u_R = M^2_{diag}$
with $M^2_{diag}=diag\{m_u^2,m_c^2,m_t^2\}$, and we find it possible
that $(M_{u})_{ct}$ is several times larger than $(M_{u})_{ut}$.

In the third-generation enhanced left-right model, beside the
dominant QCD contribution $ q \bar{q} \to g^\ast \to t \bar{t}$,
diagrams shown in Fig.\ref{fig3} also contribute to $t\bar{t}$
production at the Tevatron, and the amplitudes of these new
contributions are \small
\begin{eqnarray}
M_a &=& i\delta_{\alpha\beta}\delta_{\rho\sigma}\left(\frac{e}{2c_w
s_w}\right)^{2}
\frac{\bar{u}(t)\gamma_{\mu}[g_{ZL}^{t}P_L+g_{ZR}^{t}P_R]
v(t)\bar{v}(u)\gamma^{\mu}[g_{ZL}^{u}P_L+g_{ZR}^{u}P_R]u(u)}{(p_1+p_2)^2-m_{Z}^{2}}, \\
M_b &=& i\delta_{\alpha\beta}\delta_{\rho\sigma}\left(\frac{e}{2c_w
s_w}\right)^{2}
\frac{\bar{u}(t)\gamma_{\mu}[g_{Z^{'}L}^{t}P_L+g_{Z^{'}R}^{t}P_R]
v(t)\bar{v}(u)\gamma^{\mu}[g_{Z^{'}L}^{u}P_L+g_{Z^{'}R}^{u}P_R]u(u)}{(p_1+p_2)^2-m_{Z^{'}}^{2}-i\Gamma_{Z^{'}}m_{Z^{'}}}, \\
M_c &=& i\delta_{\alpha\rho}\delta_{\beta\sigma}\left(\frac{e}{2c_w
s_w}\right)^{2}\left[\xi_Z s_w
(\cot{\theta_R}+\tan{\theta_R})\right]^{2}|V_{Rtu}^{u}V_{Rtt}^{u}|^{2}
\frac{\bar{u}(t)\gamma_{\mu}P_R u(u)\bar{v}(u)\gamma^{\mu}P_R v(t)}{(p_1-p_3)^2-m_{Z}^{2}}, \\
M_d &=& i\delta_{\alpha\rho}\delta_{\beta\sigma}\left(\frac{e}{2c_w
s_w}\right)^{2}\left[s_w
(\cot{\theta_R}+\tan{\theta_R})\right]^{2}|V_{Rtu}^{u}V_{Rtt}^{u}|^{2}
\frac{\bar{u}(t)\gamma_{\mu}P_R u(u)\bar{v}(u)\gamma^{\mu}P_R
v(t)}{(p_1-p_3)^2-m_{Z^{'}}^{2}},
\end{eqnarray}
\normalsize where $\Gamma_{Z^\prime}$ is the $Z^\prime$ width
obtained by adding all decay channels of $Z^\prime$,
$s_w=\sin{\theta_{w}}$, $c_w=\cos{\theta_{w}}$, and the coupling
coefficients $g_{ZL}^{t,u}$, $g_{ZR}^{t,u}$, $g_{Z^{'}L}^{t,u}$ and
$ g_{Z^{'}R}^{t,u}$ are defined as
\begin{eqnarray}
g_{ZL}^{t,u} &=& 1- \frac{4}{3} s_{w}^{2} - \frac{1}{3} s_w \tan{\theta_R} \xi_Z, \\
g_{ZR}^{u} &=& -\frac{4}{3} s_{w}^{2} - \frac{4}{3} s_w \tan{\theta_R} \xi_Z, \\
g_{ZR}^{t} &=& -\frac{4}{3} s_{w}^{2} - \frac{1}{3} s_w \tan{\theta_R} \xi_Z+ s_w \cot{\theta_R} \xi_Z, \label{coup1}
\end{eqnarray}
\begin{eqnarray}
g_{Z^{'}L}^{t,u} &=& (1 - \frac{4}{3} s_{w}^{2} ) \xi_Z + \frac{1}{3} s_w \tan{\theta_R},  \label{coup2} \\
g_{Z^{'}R}^{u} &=& -\frac{4}{3} s_{w}^{2} \xi_Z + \frac{4}{3} s_w \tan{\theta_R}, \\
g_{Z^{'}R}^{t} &=& - \frac{4}{3} s_{w}^{2} \xi_Z + \frac{1}{3} s_w
\tan{\theta_R} \xi_Z - s_w \cot{\theta_R} .
\end{eqnarray}
Note that among the four amplitudes, only $M_c$ and $M_d$ interfere
with the QCD amplitude.

In this model the new parameters
$\xi_z$, $\cot{\theta}_{R}$, $M_{Z^{'}}$ and $(V^u_R)_{ut}$ are
involved in our calculation. Constraints on these parameters
were discussed in \cite{hexg}, and it was found that $ 0 \le \xi_z
\le 0.02$ and $\cot{\theta}_{R} \le 20 $. As for $M_{Z^\prime}$, we
should note that the constraints from CDF search for
$Z^\prime$\cite{zprime} and from the global fitting of the
electroweak precision data \cite{ewzprime} are invalid here since
these constraints arise mostly from the processes involving the
first- or second-generation of fermions. So far the pertinent bound
comes from $e^{+}e^{-} \to b\bar{b}$ at LEP-II, which requires
$M_{Z^\prime} \gtrsim 460$ GeV for $\cot \theta_R \ge
10$ \cite{hexg}. In following analysis, without stating explicitly,
we include the QCD contribution to $A^t_{FB}$ and scan the new free
parameters in the following ranges
\begin{eqnarray}
500 {\rm ~GeV} \le M_{Z^{\prime}} \le 2000 {\rm ~GeV},
\quad 0 \le \xi_Z \le 0.02,
\quad  10 \le \cot{\theta}_{R} \le 20, \quad 0.1 \le (V^{u}_R)_{ut}
\le 0.2 \nonumber
\end{eqnarray}

\begin{figure}[htb]
\epsfig{file=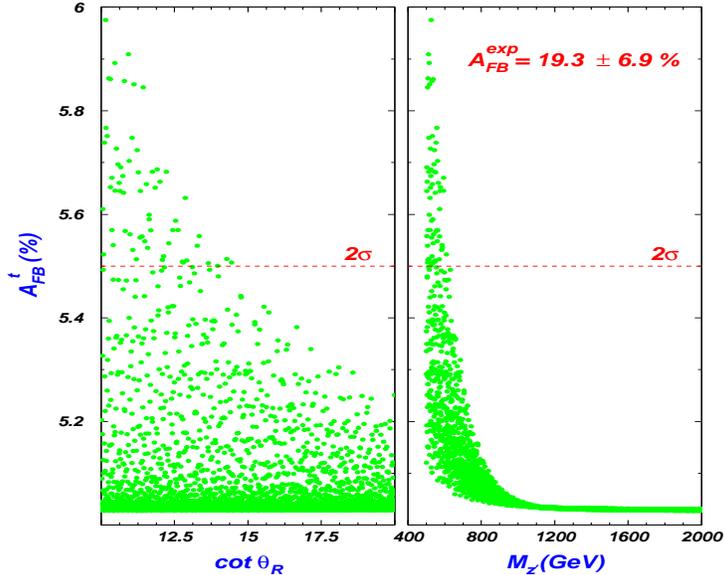,width=10cm,height=8cm}
\vspace*{-0.8cm}
\caption{Scatter plots of $A_{FB}^{t}$ versus $\cot{\theta}_{R}$ and
$m_{Z^{'}}$ for the no-mixing case in the LR model.} \label{fig4}
\end{figure}

In our discussion we consider two cases, i.e., with and without
$u_R-t_R$ mixing. For the no-mixing case, new contributions to
$A^t_{FB}$ only come from  diagrams (a) and (b) of Fig.\ref{fig3}
and the dependence of $A_{FB}^{t}$ on $\cot{\theta}_{R}$ and
$m_{Z^{'}}$ are shown in Fig.\ref{fig4}. This figure shows that a
light $Z'$ with relatively small $\cot{\theta}_{R}$ can enhance
$A_{FB}^{t}$ to the $2\sigma$ region of its experimental value. This
is because small $\cot \theta_R$ can enhance $Z^\prime \bar{u} u$
interactions, and a light $Z^\prime$ can make the resonance effect
of the diagram Fig.\ref{fig3}(b) on $t\bar{t}$ production more
significant. Fig.\ref{fig4} also shows that in the no-mixing case,
the $Z^\prime$ contribution to $A^t_{FB}$ can only reach $1\%$. This
is due to the smallness of $Z^\prime \bar{u} u$ couplings (see
Eq.(\ref{coup1}) and Eq.(\ref{coup2})) and the absence of the
interference of diagram (a) and (b) in Fig.\ref{fig3} with the
dominant QCD amplitude.

\begin{figure}[htb]
\includegraphics[width=4in,height=3.in]{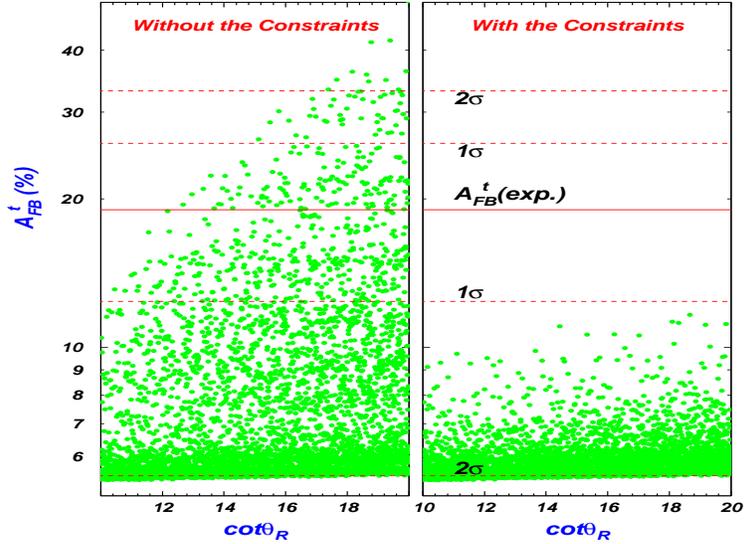}
\vspace*{-0.8cm}
\caption{Scatter plots of $A_{FB}^{t}$ versus
$\cot{\theta}_{R}$ in the mixing case of the LR model. The
left (right) panel is without (with) considering the constraints
from $\sigma_{t\bar{t}}$ and $M_{t\bar{t}}$.  } \label{fig5}
\end{figure}
\begin{figure}[htb]
\includegraphics[width=4in,height=3.in]{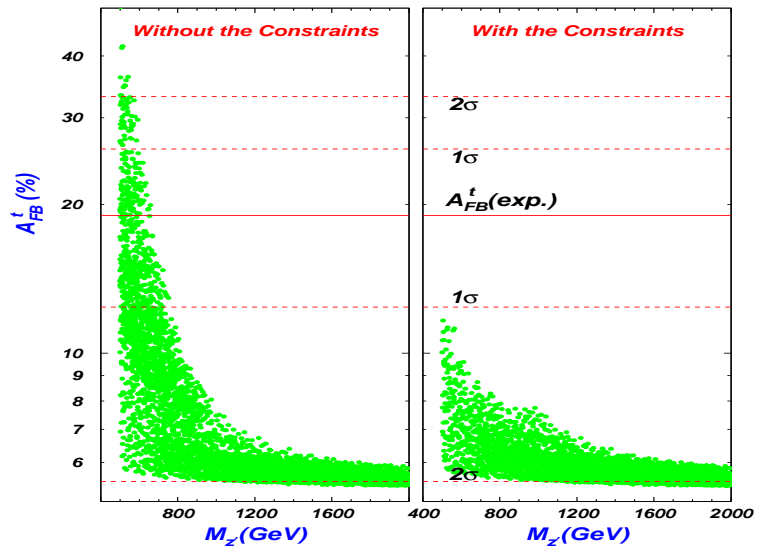}
\vspace*{-0.8cm}
\caption{Same as Fig. \ref{fig5}, but projected in
$A_{FB}^{t}$ versus $M_{Z^\prime}$ plane.  } \label{fig6}
\end{figure}
In Figs.\ref{fig5} and \ref{fig6}, we show $A^t_{FB}$ versus
$\cot \theta_R$ and $M_{Z^\prime}$ respectively in the mixing case.
The left panels of these figures show that the effect of
Fig.\ref{fig3} (mainly diagram (d)) is potentially very large,
pushing the value of $A^t_{FB} $ up to $50\%$. In this case, the
effects on the total $t \bar{t}$ production rate and the
$t\bar{t}$ invariant mass distribution are also large and
so it is necessary to consider constraints from these
observables measured at the Tevatron.
Now based on the CDF 4.6 fb$^{-1}$ luminosity data, the
measured total cross section is
$\sigma^{\rm exp}_{t \bar t} = 7.50 \pm 0.31_{stat} \pm
 0.34_{syst} \pm 0.15_{th} \ pb $ for $m_t=172.5$ GeV \cite{CDF-mt}.
Combining errors in quadrature, one can get $\sigma^{exp}_{t \bar t}
= 7.50 \pm 0.48$ pb, which is in good agreement with the SM
prediction $\sigma_{t \bar t}= 7.5^{+0.5}_{-0.7} $\ pb \cite{nlo}.
The invariant mass distribution was also measured by CDF, and the
results are presented in nine bins of $M_{t\bar{t}}$ \cite{mtt}.
When we calculate these observables, we multiply an overall K-factor
1.329 for the dominant tree-level QCD contribution \cite{nlo}. The
right panels of Figs.\ref{fig5} and \ref{fig6} are then obtained by
requiring the total cross section and the differential cross section
in each bin to be within the $2\sigma$ regions of their experimental
values. These results shows that even with the constraints, the
third-generation enhanced left-right model can still enhance
$A^t_{FB}$ to $12\%$ (well above the $2\sigma$ lower bound),
but it can not enhance $A^t_{FB}$  to within
the $1\sigma$ region of its experimental value. Note that unlike the
no-mixing case, large contribution to $A^t_{FB}$ comes from the
region where $\cot \theta_R$ is large. The reason is that in the
mixing case, dominant contribution arises from diagram (d) of
Fig.\ref{fig3}, and this contribution is proportional to $ ( \cot
\theta_R + \tan \theta_R )^2$.

\section{Discussion and Conclusion}
From our study and the previous works
\cite{top-afb-sm,top-afb-extra,top-afb-s,top-afb-t,top-afb-scalar,top-afb-ind},
we can learn that to make sizable contribution to $A^t_{FB}$, the
following two conditions must be satisfied. One is the initial state
must be $u\bar{u}$ and/or $d\bar{d}$ and in case that only
$d\bar{d}$ initiated contribution is responsible to explain
$A^t_{FB}$, the involved interaction must be strong enough to
compensate the suppression of the parton distribution of down quark
in proton. The other is the amplitude of the $t\bar{t}$ production
must contain terms proportional to $\cos \theta$ with $\theta$ being
the angle between the reconstructed top quark momentum and the
proton beam direction in $t\bar{t}$ rest frame, or in other words,
contain terms like $( p_u \cdot p_t )( p_{\bar{u}} \cdot p_{\bar{t}}
) - (p_{\bar{u}}\cdot p_t ) ( p_u \cdot p_{\bar{t}} ) $. This
requirement implies that new physics affect $t \bar{t}$ production
through the following ways:
\begin{itemize}
\item[(1)] Through $s$-channel process $q \bar q \to t \bar t$ by exchanging
a gauge boson \cite{top-afb-extra,top-afb-s}. If this process
interferes with the the $s$-channel QCD process $q \bar q\to g^*\to
t \bar t$, the interaction of  the gauge boson with $q$ and $t$ must
have axi-vectorial component. Examples in this direction are the
presence of Kaluza-Klein excitation of gluon in extra dimension
\cite{top-afb-extra} or the axigluon \cite{top-afb-s}.  If this
process does not interfere with the $s$-channel process, to affect
$A^t_{FB}$ by itself, the interaction must have both vectorial and
axi-vectorial component, like what we studied in diagram (b) of
Fig.\ref{fig3}. From yet known studies we can infer that it seems
difficult for the latter case to enhance $A^t_{FB}$ significantly
without spoiling the constraints from $\sigma_{t\bar{t}} $ and
$M_{t\bar{t}}$.
\item[(2)] Through $t$-channel process $q \bar q \to  t \bar t$ by exchanging
a vector boson or a scalar \cite{top-afb-t,top-afb-scalar}, which
interferes with the QCD process $q \bar q\to g^*\to t \bar t$. In
this way, scalar is less efficiency than vector boson in explaining
$A^t_{FB}$ given that they have the same coupling strength and mass. This is
because for the scalar case, there is a competition between spin
correlation and the Rutherfold singularity \cite{top-afb-scalar}.
Moreover, as shown in \cite{top-afb-scalar}, the scalar-mediated
contributions can be categorized by the transformation property of
the scalar under $SU(3)_c$. If the scalar is color-triplet or
-sextet, there exists large parameter region to explain $A^t_{FB}$
within $1\sigma$ and at the same time to remain $\sigma_{t\bar{t}}$
within the experimental errors, while for color-singlet or -octet
scalar, it is very difficult to produce a large positive
contribution to $A^t_{FB}$ without spoiling the constraint from
$\sigma_{t\bar{t}}$ \cite{top-afb-scalar}. In our work, we checked
this conclusion for color-singlet and -triplet cases.
\end{itemize}

From our study, we can also learn that, although the top quark pair
productions at the Tevatron and the LHC may be sensitive to new
physics, the effects of new physics (like the popular MSSM or TC2
models) are usually not so large to be well above the experimental
and theoretical uncertainties \cite{susytt-loop}. A complementary or
even more sensitive probe for new physics effects in top quark
sector is top quark FCNC processes, which are extremely suppressed
and unaccessible in the SM but can be greatly enhanced by several
orders to reach the observable level in some new physics models like
the MSSM \cite{mssm-t-fcnc} or the TC2 model \cite{tc2-top}.

In summary, we in this work calculated the new physics contribution
to the top quark forward-backward asymmetry $A_{FB}^{t}$ at the
Tevatron in two different models: the minimal supersymmetric model
without R-parity (RPV-MSSM) and the third-generation enhanced
left-right model (LR model). We found that in the parameter space
allowed by the $t\bar{t}$ production rate and the $t\bar{t}$
invariant mass distribution at the Tevatron, the LR model can
enhance $A^t_{FB}$ to within the $2\sigma$ region of the Tevatron
data for the major part of the parameter space, and in optimal case
$A^t_{FB}$ can reach $12\%$ which is slightly below the $1\sigma$
lower bound. For the RPV-MSSM, only in a narrow part of the
parameter space can the $\lambda''$ couplings enhance $A^t_{FB}$ to
within the $2\sigma$ region while the $\lambda'$ couplings just
produce negative contributions to worsen the fit. Noting that the
R-parity conserving interactions in the MSSM are unlikely to give
large enough contribution to $A_{FB}^{t}$, we conclude that the MSSM
with (without) R-parity will be disfavored (favored in case of
$\lambda''$ couplings) if the discrepancy of $A_{FB}^{t}$ persists
with more forthcoming data accumulated by the Tevatron. We also
checked the top-color assisted technicolor model and found that it
gives negligibly small contributions to $A_{FB}^{t}$ and thus
unlikely to explain the  Tevatron data.

\section*{Acknowledgement}
This work was supported in part by the National Natural Science
Foundation of China (NNSFC) under grant Nos. 10505007, 10821504,
10725526 and 10635030, by HASTIT under grant No. 2009HASTIT004 and
2010IRTSTHN002, by the Project of Knowledge Innovation Program
(PKIP) of Chinese Academy of Sciences under grant No. KJCX2.YW.W10.

\end{document}